\documentclass{ieeeaccess}
\usepackage{cite}
\usepackage{amsmath,amssymb,amsfonts}
\usepackage{algorithmic}
\usepackage{graphicx}
\usepackage{textcomp}
\usepackage{siunitx}
\usepackage{comment}
\usepackage{color}
\def\BibTeX{{\rm B\kern-.05em{\sc i\kern-.025em b}\kern-.08em
    T\kern-.1667em\lower.7ex\hbox{E}\kern-.125emX}}
\begin{document}

\title{Intranasal Chemosensory Lateralization Through the Multi-electrode Transcutaneous Electrical Nasal Bridge Stimulation}
\author{\uppercase{Ayari Matsui}\authorrefmark{1}, \uppercase{Tomohiro Amemiya}\authorrefmark{2,3}, \uppercase{Takuji Narumi}\authorrefmark{1}, \uppercase{Kazuma Aoyama}\authorrefmark{2,4}}
\address[1]{Graduate School of Information Science and Technology, The University of Tokyo, Hongo, Bunkyo-ku, Tokyo 113-8656, Japan}
\address[2]{Virtual Reality Educational Research Center, The University of Tokyo, Hongo, Bunkyo-ku, Tokyo 113-8656, Japan}
\address[3]{Information Technology Center, The University of Tokyo, Yayoi, Bunkyo-ku, Tokyo 113-8658, Japan}
\address[4]{Faculty of Informatics, Gunma University, Aramaki-cho, Maebashi, Gunma 371-8510, Japan}

\tfootnote{This work involved human participants or animals in its research. The University of Tokyo approved all ethical and experimental procedures and protocols.}

\markboth
{Matsui \headeretal: Intranasal Chemosensory Lateralization Through the Multi-electrode Transcutaneous Electrical Nasal Bridge Stimulation}
{Matsui \headeretal: Intranasal Chemosensory Lateralization Through the Multi-electrode Transcutaneous Electrical Nasal Bridge Stimulation}

\corresp{Corresponding author: Ayari Matsui (e-mail: matsui@cyber.t.u-tokyo.ac.jp)}

\begin{abstract}
Numerous studies have been conducted on display techniques for intranasal chemosensory perception. However, a limited number of studies have focused on the presentation of sensory spatial information.
To artificially produce intranasal chemosensory spatial perception, we focused on a technique to induce intranasal chemosensation by transcutaneous electrical stimulation between the nasal bridge and the back of the neck.
Whether this technique stimulates the trigeminal nerve or the olfactory nerve remains debatable; if this method stimulates the trigeminal nerve, the differences in the amount of stimulation to the left and right trigeminal branches would evoke lateralization of intranasal chemosensory perception.
Therefore, we propose a novel method to lateralize intranasal chemosensation by selectively stimulating the left or right trigeminal nerve branches through the shifting of an electrode on the nasal bridge to the left or right. 
Finite element simulations reveal that electrical stimulation applied between the electrodes on the left/right nasal bridge and the back of the neck results in the construction of a high current density area on the left/right branch of the trigeminal nerve. 
The results of two psychophysical experiments reveal that intranasal chemosensation can be lateralized by using the proposed method.
The results of our experiment also suggest that lateralization is not the result of electrically induced tactile sensation of the skin surface but rather due to the distribution of stimuli to the trigeminal nerves.
To the best of our knowledge, this study is the first successful lateralization of intranasal chemosensation that utilizes an easy-to-apply method without involving nostril blocking.
\end{abstract}


\titlepgskip=-15pt

\maketitle

\section{Introduction}
\label{sec:introduction}
Technologies that render intranasal chemosensation, including olfaction and trigeminal sensation, have attracted considerable research attention because of their potential for enhancing the sense of immersion in virtual reality (VR).
For example, Archer \textit{et al.} reported that adding smell to VR games considerably improved the sense of immersion and the psychological and physiological quality of the VR experience\cite{Archer2022}.
Various methods, including devices that deliver chemical substances to the nose through micro dispensers and solenoid valves\cite{Nakamoto2018} and methods that induce intranasal chemosensation through electrical stimulation\cite{Hariri2016, Holbrook2019}, have been proposed to provide chemosensations.

However, most intranasal chemosensory studies have focused on inducing sensation, but limited studies have focused on the spatial rendering of intranasal chemosensation.
Intranasal chemosensory displays that can accurately represent the spatial information of chemosensations could be used for the precise representation of intranasal chemosensation in VR, which can result in immersive and realistic experiences.
These intranasal chemosensory displays also allow applications such as guidance using smells.

Regarding the spatial perception of intranasal chemosensation, the stimulus distribution to the trigeminal nerve branches in the nasal cavity is thought to be essential\cite{Kobal1989, Frasnelli2008, Kleemann2009, Lundstrom2012, Croy2014}.
Olfactory and trigeminal nerve bundles exist in the nasal cavity, and intranasal chemosensation results from the interaction of these inputs\cite{Hummel2002}.
Olfactory nerves detect odorants and induce olfaction, whereas trigeminal nerves detect pungent gases, such as ammonia and carbon dioxide (CO$_2$), and evoke stinging sensations.
Olfactory and trigeminal inputs are combined and perceived as intranasal chemosensation, whereas the trigeminal input alone is responsible for the spatial perception of sensory information.
The spatial perception of intranasal chemosensation, particularly lateralization, has been investigated in numerous experiments by using pure olfactory stimuli, olfactory-trigeminal stimuli, and pure trigeminal stimuli.
The results of these experiments have revealed that the trigeminal input is necessary for lateralizing sensations\cite{Kobal1989, Frasnelli2008, Kleemann2009, Lundstrom2012, Croy2014}.
For example, Kleemann \textit{et al.} reported that participants could lateralize the trigeminal stimulants isoamyl acetate and CO$_2$ with high probability, whereas they could not lateralize hydrogen sulfide, which is a pure odorant that does not stimulate the trigeminal nerve\cite{Kleemann2009}.
Another study has reported that the ability of lateralization in olfaction can be improved through training\cite{Negoias2013}. 
However, olfactory fMRI scanning of individuals with high lateralizing ability reveals a significantly enhanced activation of cerebral trigeminal processing area.
This suggests that the lateralization of olfaction is determined by the degree to which olfactory input activates the trigeminal area; moreover, this activation level depends on the type and intensity of the olfactory input, as well as individual differences, and can potentially be improved through training\cite{Croy2014}. 
Therefore, trigeminal activation evidently plays an important role in the lateralization of intranasal chemosensation.
Several studies have indicated that humans can detect the direction of intranasal chemosensation without a trigeminal input at the subconscious level.
Wu \textit{et al.} reported that even with nontrigeminal odor stimuli, the self-motion direction of the participants is biased toward the high concentration side\cite{Wu2020}.
However, they confirmed that the participants could not verbalize whether they perceive the odor in the left or right nostril without trigeminal input.
Thus, the trigeminal input is necessary to clearly distinguish the left and right sides of chemoperception, at least on verbally recognizable levels.
On the basis of these studies, the distribution of input stimuli to the left and right trigeminal nerves lateralizes intranasal chemosensation.

Several methods of trigeminal nerve stimulation, such as chemical stimulation using CO$_2$, electrical stimulation, and mechanical stimulation using an air puff, have been proposed \cite{Iannilli2008}.
Among these methods, electrical stimulation is the most compact and inexpensive; moreover, the timing and intensity of stimulation can be easily controlled through computers.

Various studies have used electrical stimulation to activate the olfactory and trigeminal systems. Maharjan \textit{et al.} reported that high-frequency (80 Hz) stimulation of the auricular branch of the vagus nerve can enhance olfactory performance, while Badran \textit{et al.} demonstrated that electrical stimulation of the trigeminal nerve through forehead electrodes can improve olfactory performance\cite{Maharjan2018, Badran2022}. 
Cakmak \textit{et al.} used simulations to suggest that electrodes placed on the nasal bridge and the back of the neck can activate the olfactory region, but experiments with human subjects have not been conducted to support their study\cite{Cakmak2020}.

Aoyama \textit{et al.} discovered that applying direct current between electrodes across the nasal bridge and on the back of the neck can induce intranasal irritating chemosensation, which they named galvanic olfactory stimulation (GOS)\cite{Aoyama2021}.
Although whether GOS stimulates the olfactory nerve or the trigeminal nerve remains debatable, the features of the sensation reported by the participants indicated that it likely stimulates the trigeminal nerve.
If GOS acts on the trigeminal nerve, left--right differences in stimulus intensity evoke the lateralization of intranasal chemosensation.

The current density distribution on the nerves produced by electrical stimulation can be changed by the position of the electrodes.
Therefore, various electrode positions generate a biased current density on the left and right trigeminal nerves.
According to previous studies, we hypothesized that the generation of biased current densities on both sides of trigeminal nerves enables the lateralization of intranasal chemosensory perception.
This study revealed that by shifting the electrode position on the nasal bridge of the GOS, which was proposed in the previous study\cite{Aoyama2021}, to the left or right, the trigeminal nerve on the left or right side was selectively stimulated, which allowed control of the position where intranasal chemosensation arises.

To the best of our knowledge, this study is the first that lateralizes intranasal chemosensation through electrical stimulation from outside the nasal cavity.
This study enables the spatial rendering of intranasal chemosensation, which is yet to be investigated comprehensively.
The results of the study could be used in VR and human--computer interaction (HCI) for immersive and higher-quality physiological and psychological experiences and applications such as guidance through chemosensory directions.

\subsection{Related Work}
Numerous studies have been conducted on intranasal chemosensory displays through electrical stimulation.
However, many methods require surgical operations to attach the device or are uncomfortable to wear and are not suitable for practical use such as VR or HCI.
For example, Kumar \textit{et al.} reported that stimulation to the olfactory bulb and olfactory tract with subdural electrodes in epileptic patients resulted in most patients being able to smell\cite{kumar2012}.
Holbrook \textit{et al.} placed electrodes in the ethmoid sinus and applied a stimulating current to participants who had undergone sinus surgery, including total ethmoidectomy, and found such measures induced unpleasant odors \cite{Holbrook2019}.
These invasive techniques require surgery and cannot be easily used by everyone.
Considering applications such as VR with the spatial representation of intranasal chemosensation and navigation through chemosensations such as smells, the method that does not require surgery and is comfortable for users to wear is ideal.
As a minimally invasive technique that does not require surgery, Hariri \textit{et al.} proposed a novel method of electrical stimulation to the olfactory region by inserting electrodes deep into the nasal cavity\cite{Hariri2016}.
Brooks \textit{et al.} specifically targeted the trigeminal input to intranasal chemosensation.
They placed electrodes across the nasal septum and reported that the absolute electric charge of the stimulation modulates the intensity of the induced intranasal chemosensation, whereas the phase order and net charge determine the direction of the sensation\cite{Brooks2021}.
However, even though these methods are minimally invasive, 
they are not comfortable to wear as electrodes are inserted into the nostrils, and chemical olfactory stimulation cannot be performed simultaneously as they partially block the nostrils.

The electrical stimulation from the outside of the nasal cavity by placing electrodes on the nasal bridge and on the back of the neck as proposed by Aoyama \textit{et al.} successfully induced intranasal chemosensation less invasively and without any obstruction in the nostrils\cite{Aoyama2021}.
This study followed Aoyama \textit{et al.} in placing electrodes on the nasal bridge and on the back of the neck and shifted the electrode on the nasal bridge from the center to left or right to stimulate the left or right trigeminal nerve selectively.

Nakamura \textit{et al.} demonstrated that selective stimulation of nerves is possible depending on the position of electrodes\cite{Nakamura2021} and developed a multi-electrode galvanic taste stimulation (GTS) configuration with electrodes outside the mouth. 
The results of their study revealed that the electrical potential distribution within the buccal cavity could be manipulated, which could be used to control where taste sensation is induced.
According to this study, the electrical current density distribution in the nasal cavity can theoretically be controlled by using multiple electrodes on the nasal bridge and thus can modulate the location of chemosensation.

The novelty of this study is that it succeeds in manipulating the location of intranasal chemosensation with a minimally invasive and easy-to-wear method. 
Psychophysical experiments confirmed that chemosensory lateralization is not a perceptual illusion caused by the tactile sensation of electrical stimulation to the skin.
Furthermore, our experimental results, in which the lateralization of intranasal chemosensation was observed, provide scientific novelty in which the sensations induced by electrical stimulation between electrodes on the nasal bridge and the back of the neck can be attributed to the trigeminal nerve input.

\subsection{Stimulation and Study Design}
This study developed a simple and easy-to-wear method to control the location of intranasal chemosensory perceptions.
To induce intranasal chemosensation, the method proposed by Aoyama \textit{et al.}\cite{Aoyama2021} was used.
In this method, electrodes are placed on the nasal bridge and the back of the neck, and an electric current was applied between these electrodes.
In our study, we call this method transcutaneous electrical nasal bridge stimulation (TENS).

To control the location of the intranasal chemosensation evoked by the TENS, two types of TENS were examined in this study: the left TENS and the right TENS.
In the left TENS, the electrode of the nasal bridge was shifted to the left side compared with the center TENS, in which the electrode was located in the center of the nasal bridge.
Similarly, in the right TENS, the electrode on the nasal bridge was attached to the right side.
Based on previous studies, the hypothesis was that the left TENS stimulates the left trigeminal branch strongly and vice versa, thus lateralizing the intranasal chemosensation whose spatial perception is mediated by the distribution of trigeminal stimulations.

To test this hypothesis, a simulation for these stimulation designs was conducted, followed by two psychophysical experiments.

\section{Simulation for the Electrode Configuration}
\begin{figure}[htbp]
 \centering
 \includegraphics[width=0.9\columnwidth]{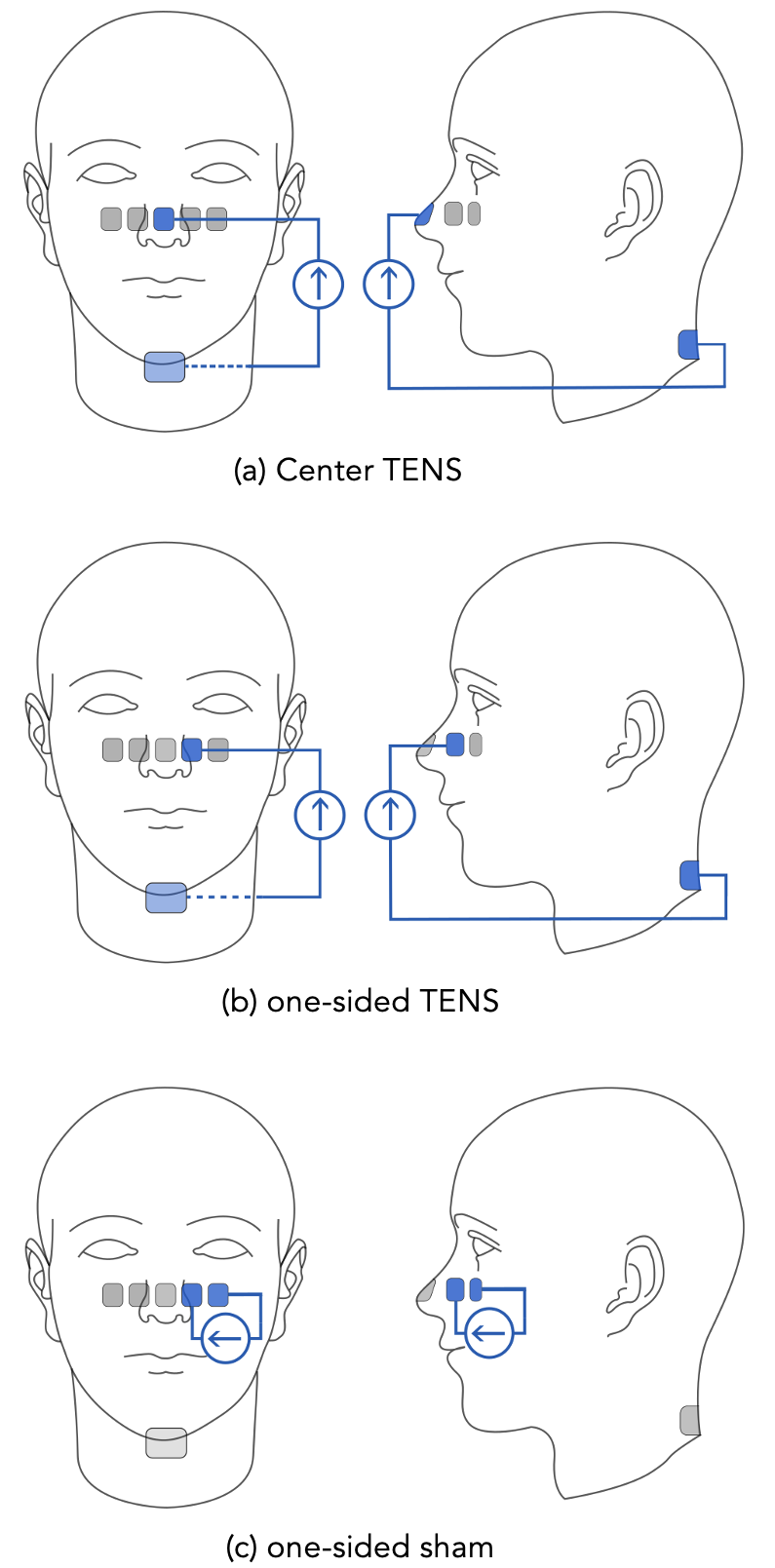}
 \caption{Positions of the electrodes. (a) Center TENS, (b) one-sided TENS, left TENS in this figure, and (c) one-sided sham, the left sham in this figure. In the right TENS and the right sham, the current is applied between the right nasal bridge and the back of the neck and between the right nasal bridge and the right cheek, which is contralateral to the electrode positions in (b) and (c).}
 \label{fig:electrodes}
\end{figure}

In this study, electrodes were placed on the nasal bridge and the back of the neck to stimulate the peripheral trigeminal nerve.
To investigate the relationship between the position of the electrodes on the nasal bridge and the location of induced intranasal chemosensation, three TENS stimulation configurations were designed: the left TENS, the center TENS, and the right TENS.
Electrodes were placed on the center of the nasal bridge in the center TENS as displayed in Fig. \ref{fig:electrodes}(a) and on the left side in the left TENS and vice versa in the right TENS as displayed in Fig. \ref{fig:electrodes}(b).

In addition to TENS, sham stimuli were designed for psychophysical experiments.
In the sham stimulus, electrodes were placed on the left or right side of the nasal bridge and the upper part of the cheek as displayed in Fig. \ref{fig:electrodes}(c), and an electric current was applied between them.
The current between the nasal bridge and the cheek provides tactile sensation from electrical stimulation to the nasal surface, but the strength of the stimulation to the trigeminal nerve should be less than TENS stimulation.
Therefore, sham stimulation is applied to the contralateral side while stimulating the left or right TENS to verify that the lateralization of intranasal chemosensory perception is not a perceptual illusion caused by the tactile sensation of electrical stimulation.

This study designed left/right TENS, but it is uncertain whether these TENSs cause a bias in the amount of stimulation to the left and right trigeminal branches.
It is also unclear whether the sham stimulation results in less trigeminal input compared with the two TENSs.
Therefore, to confirm the effects of the TENSs and sham stimulation, simulations were conducted to visualize the current density distribution using the finite element analysis.

\subsection{Numerical Anatomical Model}
\begin{figure*}[htbp]
 \centering
 \includegraphics[width=0.8\linewidth]{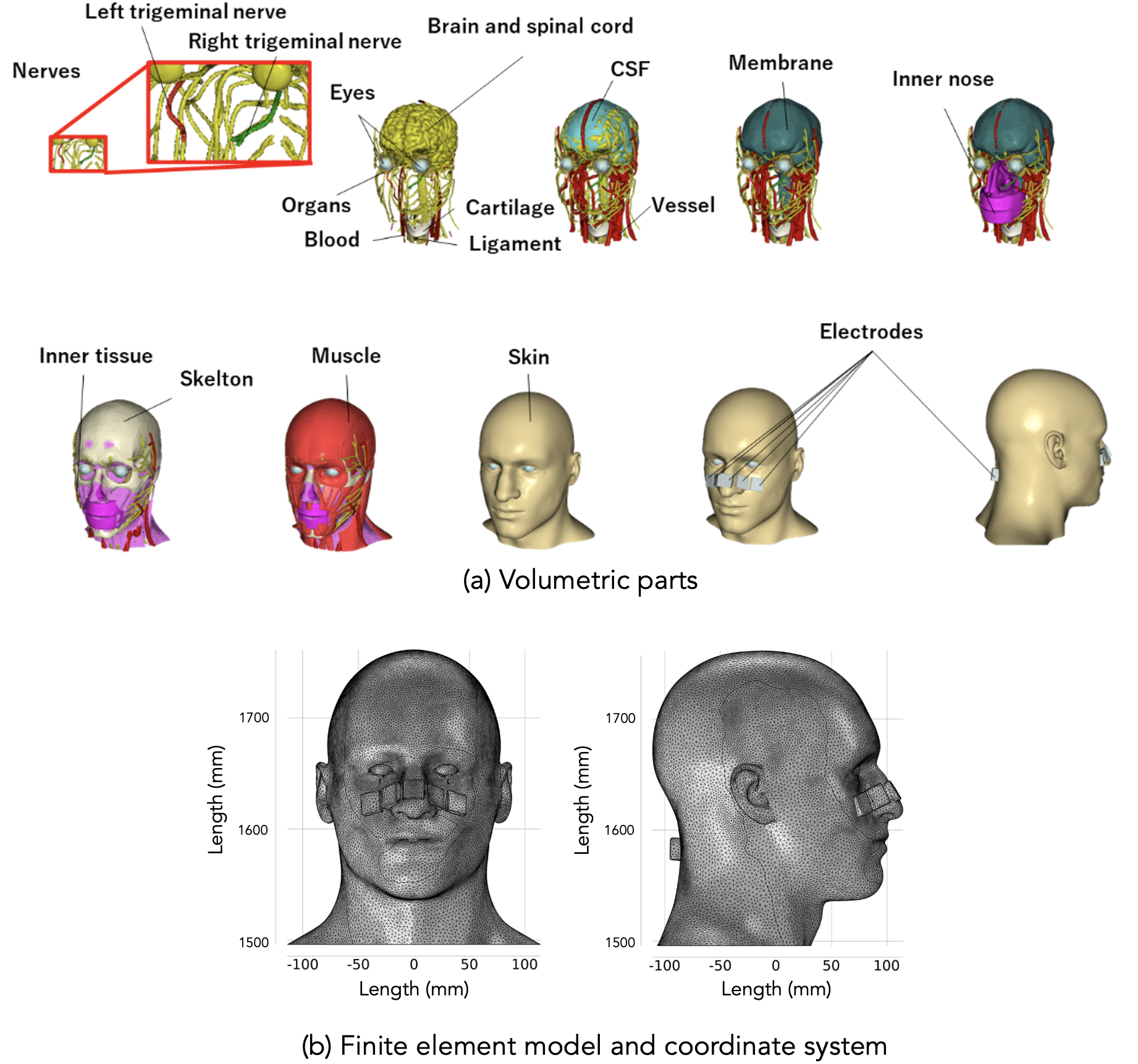}
 \caption{Head model and positions of electrodes. (a) Volumetric parts and (b) finite element model and coordinate system.}
 \label{fig:simulationHeadModel}
\end{figure*}

As the volumetric conductor model of the human body, a 3D solid male model (Zygote Media Group Inc.) was used. The model is based on data obtained from magnetic resonance imaging (MRI) and computed tomography (CT) as well as anatomical knowledge. The model is widely used in anatomical educational contexts.
The model consists of over 1,200 files (.iges or .igs format) of the entire body. 
After importing this model into Scan IP (Simpleware, SYNOPSYS Inc.), the data were grouped into 12 structures, namely brain and spinal cord, nerves, membranes, ligaments, cartilage, vessels, eyes, inner nose, organs, skeleton, muscles, and skin.
After grouping, the bottom part below the neck was removed from the entire body model. 
This model had some unnatural gaps because it did not contain internal tissues, cerebrospinal fluid (CSF), or blood. 
Therefore, the gaps were filled using the Boolean operation function of Scan IP, and the filled areas between the brain and the membranes were defined as CSF, inner vessels as blood, and all other areas as inner tissue. 
Electrodes were attached to the left side, center, and right side of the nasal bridge, the back of the neck, and the left and right cheeks. 
A \SI{20}{mm} x \SI{20}{mm} x \SI{15}{mm} rectangular electrode model was used to simulate the actual electrode area used in the later psychophysical experiments.
Owning to the difficulty of modeling an electrode that precisely fits the curvature of the skin, a thickness of \SI{15}{mm} was set to ensure that the electrode-to-skin contact area approximated that used in subsequent psychophysical experiments.
A total of 16 parts were defined in the model, as displayed in Fig. \ref{fig:simulationHeadModel}(a).

\subsection{Simulation of the current density distribution on the head}
The model edited in Scan IP was exported as a volumetric mesh file in NASTRAN format (total of 4,067,250 elements), and the file was imported into COMSOL Multiphysics 6.0 (COMSOL Inc.), as displayed in Fig. \ref{fig:simulationHeadModel}(b). 
Because COMSOL Multiphysics requires the conductivity of each part to be defined, conductivities were assigned as presented in Table \ref{tab:CurrentIDistribution}.
These conductivities were defined based on the database of IT'IS Foundation\cite{Hasgall2018} and a previous study that simulated transcranial direct current stimulation\cite{Datta2009}.

The Laplace equation $\nabla \cdot (\sigma \nabla V) = 0$ ($V$ is electrical potential, $\sigma$ is conductivity) was solved by applying the following boundary conditions:
\begin{enumerate}
   \item Inward current = $Jn$ (normal current density) is applied to the exposed surface of the anode.
   \item Ground (cathode) is applied to the exposed surface of the cathode.
   \item All other external surfaces are treated as insulators.
   \item The inward current density for each electrode is defined as appropriate to provide a current value of \SI{2.0}{mA}, considering the surface size of each electrode. 
\end{enumerate}
Simulations were conducted for a total of four conditions, namely left TENS, right TENS, left sham, and right sham.

\begin{table}[]
\caption{Electrical conductivity of tissues and body fluids.}
\label{tab:CurrentIDistribution}
    \begin{tabular}{ll}
    \hline
    Tissue                     & Conductivity (\SI{}{S/m}) \\ \hline
    Electrodes                 & 0.3                \\
    Inner tissue               & 0.465              \\
    Skin                       & 0.465              \\
    Blood                      & 0.7                \\
    Vessel                     & 0.25               \\
    Organs                     & 0.465              \\
    Muscle                     & 0.2                \\
    Membranes                       & 0.5                \\
    Cerebro spinal fluid (CSF) & 2.0                \\
    Ligaments                  & 0.25               \\
    Eye                        & 1.5                \\
    Cartilage                  & 0.15               \\
    Skeleton                      & 0.02               \\
    Brain and spinal cord      & 0.04               \\
    Inner Nose                 & 0.25               \\
    Nerves                     & 0.006             
    \end{tabular}
\end{table}

\subsection{Simulation Results}
\label{subsection:SimulationResults}
\begin{figure*}[htbp]
 \centering
 \includegraphics[width=0.9\linewidth]{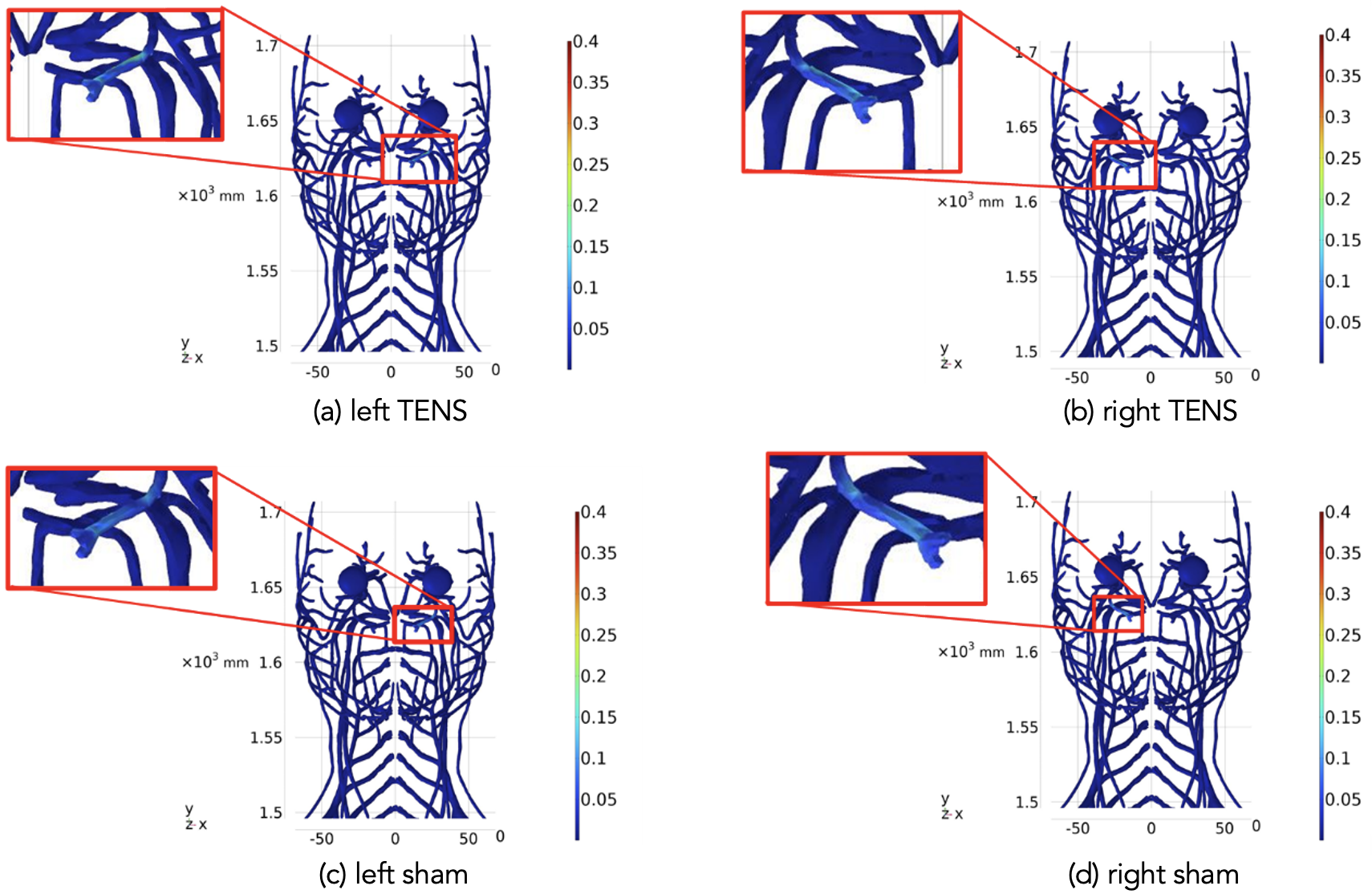}
 \caption{Simulation results of the current density distribution on the volumetric model of nerves for (a) left TENS, (b) right TENS, (c) left sham, and (d) right sham condition. Red indicates \SI{0.4}{A/m^2} or higher and dark blue indicates \SI{0}{A/m^2}. Red framed windows reveal the zoom of the maxillary nerve, which is a branch of the trigeminal nerve.}
 \label{fig:currentDensityDistribution}
\end{figure*}

\begin{figure}[htbp]
 \centering
 \includegraphics[width=\columnwidth]{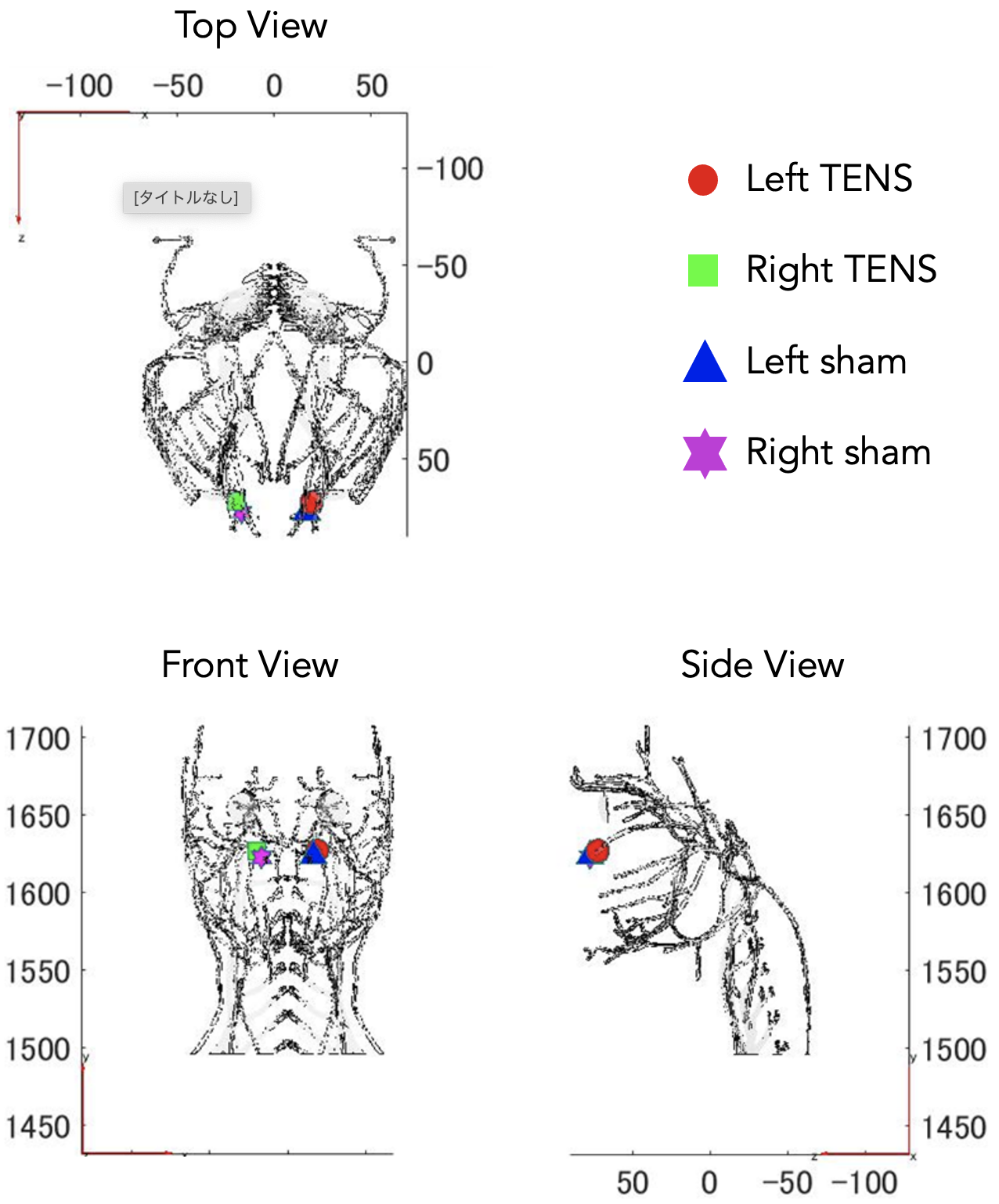}
 \caption{Points having the maximum current density on the nerves for each condition.}
 \label{fig:maximumCurrentDensityPoints}
\end{figure}

Fig. \ref{fig:currentDensityDistribution} displays the current density distribution on the volumetric model of nerves.
Red indicates that the current density is high, and the color is the darkest at \SI{0.4}{A/m^2} or higher.
Blue indicates a low current density, and the deepest blue represents \SI{0}{A/m^2}.
Table \ref{tab:maximumCurrentDensity} details the coordinates of the point of maximum current density and its current density, and Fig. \ref{fig:maximumCurrentDensityPoints} depicts the position of the maximum current density point on the nerves.

\begin{table}[]
\caption{Maximum current density points and corresponding current densities for each condition.}
\label{tab:maximumCurrentDensity}
\begin{tabular}{ccccc}
           & \multicolumn{4}{c}{Maximum Current Density}                                                                 \\ \cline{2-5} 
           & \multicolumn{3}{c}{Coordinates of the maximum point}                  & \multicolumn{1}{c}{Current density} \\ \cline{2-5} 
           & \multicolumn{1}{c}{x} & \multicolumn{1}{c}{y} & \multicolumn{1}{c}{z} & \multicolumn{1}{c}{\SI{}{A / m^2}}          \\ \hline
Left TENS   & 19.22                 & 1627                & 72.21                & 5.4315                             \\
Right TENS  & -19.42                 & 1626.9                & 72.13                 & 5.3312                             \\
Left sham  & 16.71                & 1623.5                & 72.30                & 3.2709                             \\
Right sham & -16.87               & 1622.7                & 77.34                & 3.3221                            
\end{tabular}
\end{table}

Figs. \ref{fig:currentDensityDistribution}(a) and (b) reveal that the left TENS has a high current density in the left maxillary nerve, which is a trigeminal nerve branch, and the right TENS has a high current density in the right maxillary nerve.
Thus, the left and right trigeminal nerves could be selectively stimulated by left and right TENS configurations.

As displayed in Figs. \ref{fig:currentDensityDistribution}(c) and (d), the trigeminal nerve has a slightly higher current density in the sham stimulation, but to a lesser extent than that in the TENS.
Table \ref{tab:maximumCurrentDensity} and Fig. \ref{fig:maximumCurrentDensityPoints} reveal that although the maximum current density points are positioned nearly the same in the TENS and the sham stimulation, the maximum current density in the sham stimulation is \SI{3.2709}{A/m^2} for the left sham and \SI{3.3221}{A/m^2} for the right sham, which is considerably smaller than those of TENSs.
Therefore, stimulation to the trigeminal nerve is weaker with sham stimulation than with the TENS.

Since these stimuli are expected to stimulate not only the maxillary nerve but also the ophthalmic nerve, which is closely located to the maxillary nerve, the maximum current density was also calculated for the ophthalmic nerve (Table \ref{tab:maximumCurrentDensityOnOphthalmicNerve}).

\begin{table}[]
\caption{Maximum current density on the ophthalmic nerve for each condition.}
\label{tab:maximumCurrentDensityOnOphthalmicNerve}
\begin{tabular}{ccc}
           & \multicolumn{2}{c}{Maximum Current Density (\SI{}{A / m^2})} \\ \cline{2-3} 
           & Left ophthalmic nerve                          & Right ophthalmic nerve                          \\ \hline
Left TENS   & 1.3641                                         & 0.7795                                          \\
Right TENS  & 0.8152                                         & 1.3777                                          \\
Left sham  & 1.0462                                         & 0.3700                                          \\
Right sham & 0.4215                                         & 1.1542                                         
\end{tabular}
\end{table}

Based on the simulation results, the left TENS stimulates the left trigeminal branch, whereas the right TENS stimulates the right trigeminal branch.
Thus, the left TENS would induce intranasal chemosensation on the left side of the nasal cavity and vice versa.
Furthermore, the stimulus level of the sham stimulation to the trigeminal nerve is weaker than that of the TENS.
Thus, the sham stimulation could induce some intranasal chemosensory perception, which is considerably weaker than that of TENSs.

The level of stimulation to the ophthalmic nerve by the TENS and the sham stimulation was weak compared to that to the maxillary nerve.
Therefore, although these stimulations could trigger phosphenes, the extent of the induced phosphenes is limited.
In addition, to eliminate the influence of phosphenes on the lateralization of intranasal chemosensory perception, an experimental design in which the sham stimulation is provided simultaneously on the side contralateral to the TENS can virtually eliminate the influence of left--right differences in ophthalmic nerve stimulation since the maximum current densities for the left and right ophthalmic nerves are approximately the same in the left TENS + right sham and right TENS + left sham conditions.

The first psychophysical experiment was conducted to investigate whether the position of the electrode on the nasal bridge of the TENS modulates the perceived location of intranasal chemosensation.
Another psychophysical experiment was designed using the sham stimulation to the contralateral side of the TENS simultaneously to verify that the lateralization of intranasal chemosensation is neither because of the tactile stimulation to the skin nor the phosphenes but rather the distribution of trigeminal stimulations.

\section{Psychophysical Experiments}
Ten participants (sex: 9 male, 1 female; mean age: 28.8 years; standard deviation (SD): 5.2 years) participated in the first and the second experiments.
The experiments were approved by the ethics committee at the University of Tokyo.
All participants were informed of the experiments and signed a letter of consent stating that 1) they
sufficiently understood the experimental procedures, including the risks of electrical stimulation; 2) they consented to the use of data from these experiments for publication of academic papers; and 3) they participated in these experiments voluntarily.
The study protocol was performed in accordance with the ethical standards outlined in the Helsinki Declaration.

\subsection{First Experiment: Investigation of Chemosensation Arising Area }
This experiment investigated the relationship between the position of the cathodes on the nasal bridge and the location of intranasal chemosensation.

\begin{figure}[htbp]
 \centering
 \includegraphics[width=\columnwidth]{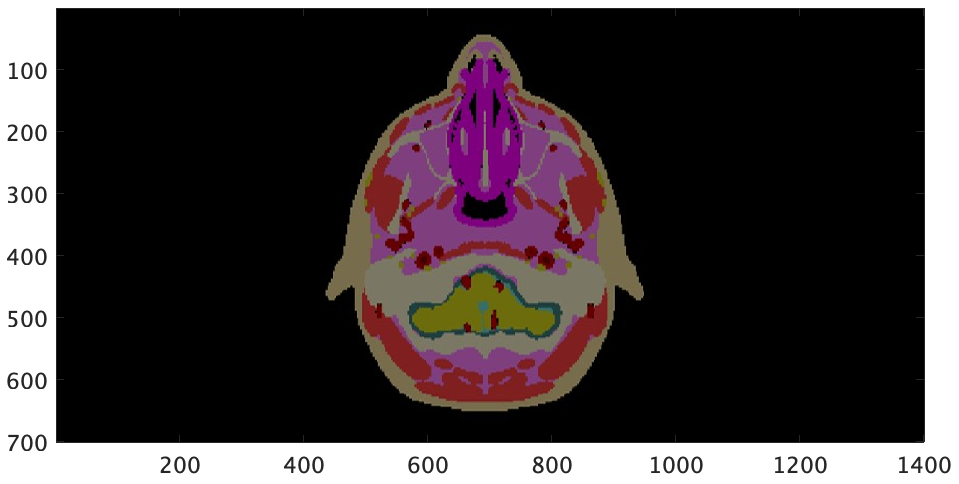}
 \caption{Cross-sectional view of the head on which the participant drew the area where they perceived intranasal chemosensation and the point of maximum intensity.}
 \label{fig:headimage}
\end{figure}

Three stimulus conditions were examined in the experiment: the left TENS, the center TENS, and the right TENS.
Although the anode was attached to the back of the neck, the cathode was attached to either the left side, the center, or the right side of the nasal bridge (Red Dot, 3M Company, on the nasal bridge and UltraStimX, Axelgaard Manufacturing Co., Ltd., on the back of the neck).
The size of electrodes was \SI{20}{mm} x \SI{20}{mm}.
A \SI{3.0}{mA}, \SI{3000}{ms} direct current was applied to the electrodes.
Participants were instructed to draw the area where they perceived intranasal chemosensation after each trial, as well as the point of most intense sensation, on the cross-sectional view of the head as displayed in Fig. \ref{fig:headimage}. 
A system developed with Processing 3 was used for the drawing.
With this system, participants used the mouse to draw, and the keyboard to change the thickness of lines and to erase lines.
Participants used this system on a laptop computer (MouseComputer Co., Ltd., mouse K5-WA, screen size 15.6 in) to draw.
Participants were told in advance to draw the region and the maximum intensity point outside the head in Fig. \ref{fig:headimage} in case they did not feel any intranasal chemosensation during a trial.
Three trials were performed for each of the three cathode position patterns to yield nine trials per participant.

\subsection{Second Experiment: Verification of the lateralization of intranasal chemosensation}
The second experiment was conducted to examine whether different positions of the cathodes on the nasal bridge enhance the lateralization of intranasal chemosensation.
Specifically, this experiment was used to verify that the lateralization of intranasal chemosensation, as indicated in Experiment 1, is not some illusory effect of the electrical tactile sensation on the skin but is caused by the distribution bias of the stimuli to the trigeminal nerve branches.

\begin{figure}[htbp]
 \centering
 \includegraphics[width=\columnwidth]{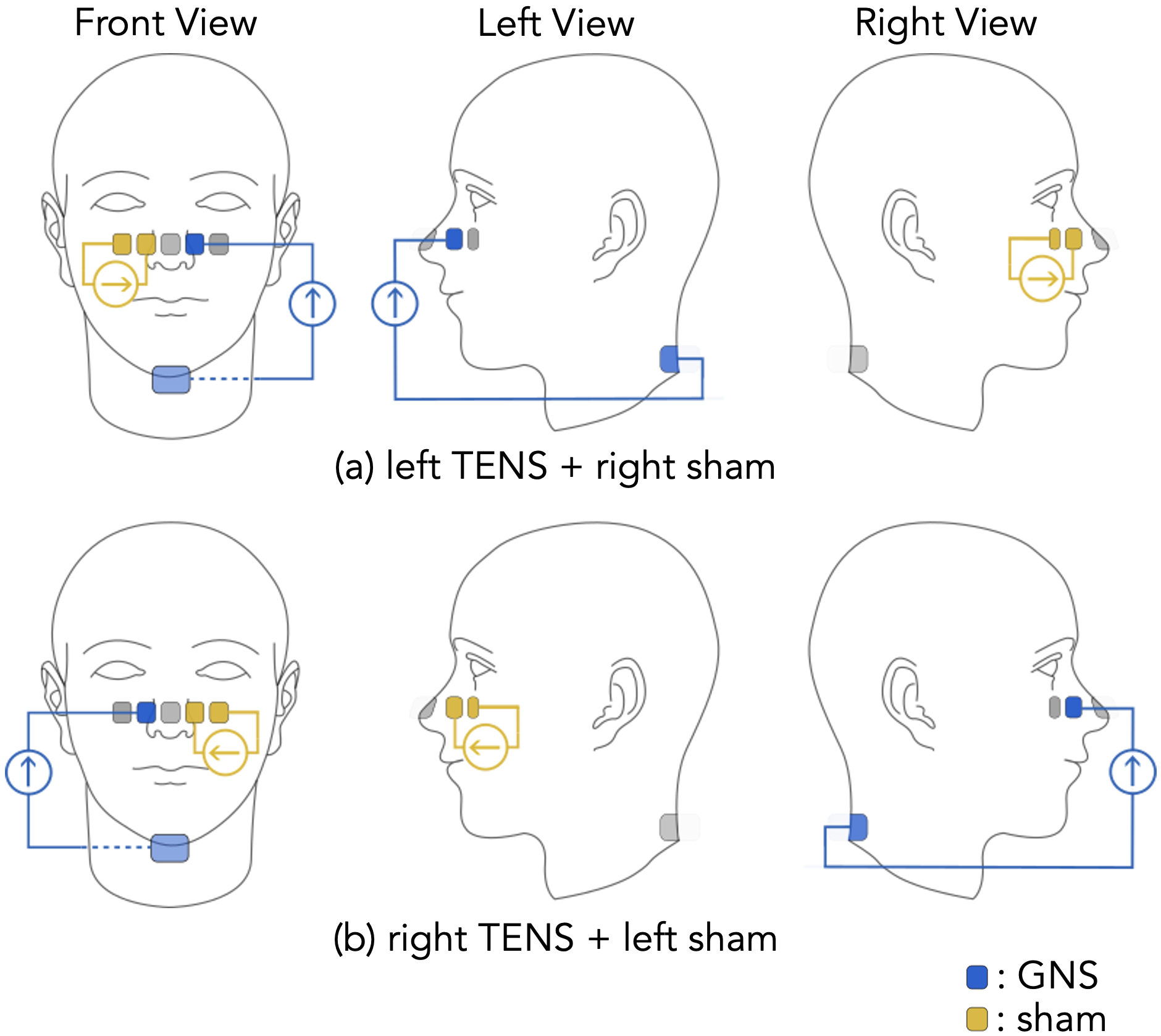}
 \caption{Two stimulus conditions examined in Experiment 2. (a) the left TENS + the right sham and (b) the right TENS + the left sham.}
 \label{fig:Ex2Stimulations}
\end{figure}

The following two stimulus conditions were examined in the experiment as displayed in Fig. \ref{fig:Ex2Stimulations}: the left TENS + the right sham and the right TENS + the left sham.
To verify that participants were not responding to skin sensations from electrical stimulation, sham stimulation was applied to the contralateral side to which the TENS was applied.
For the left TENS, the anode on the right cheek and the cathode on the right nasal bridge were sham stimulated, whereas in the right TENS condition, the anode on the left cheek and the cathode on the left nasal bridge were sham stimulated (Red Dot, 3M Company, on the nasal bridge and on the cheek, and UltraStimX, Axelgaard Manufacturing Co., Ltd., on the back of the neck).
The size of electrodes was \SI{20}{mm} x \SI{20}{mm}.
A \SI{2.0}{mA}, \SI{3000}{ms} direct current was used for both TENS and sham stimulation.
A current value of \SI{2.0}{mA} was adopted in this experiment to reduce the amount of current applied to the head for the safety of the participants.
Participants were asked to select whether they perceived the intranasal chemosensation from the left or right side of the nasal cavity by using two-alternative forced choice (2AFC).
Four trials were conducted for each condition, which yielded eight trials per participant.

\section{Results of Psychophysical Experiments}
\subsection{First Experiment: Investigation of Chemosensation Arising Area}
Prior to the analysis, data were excluded if the drawn intranasal chemosensation area and maximum intensity point were positioned outside of the head.

\begin{figure}[htbp]
 \centering
 \includegraphics[width=\columnwidth]{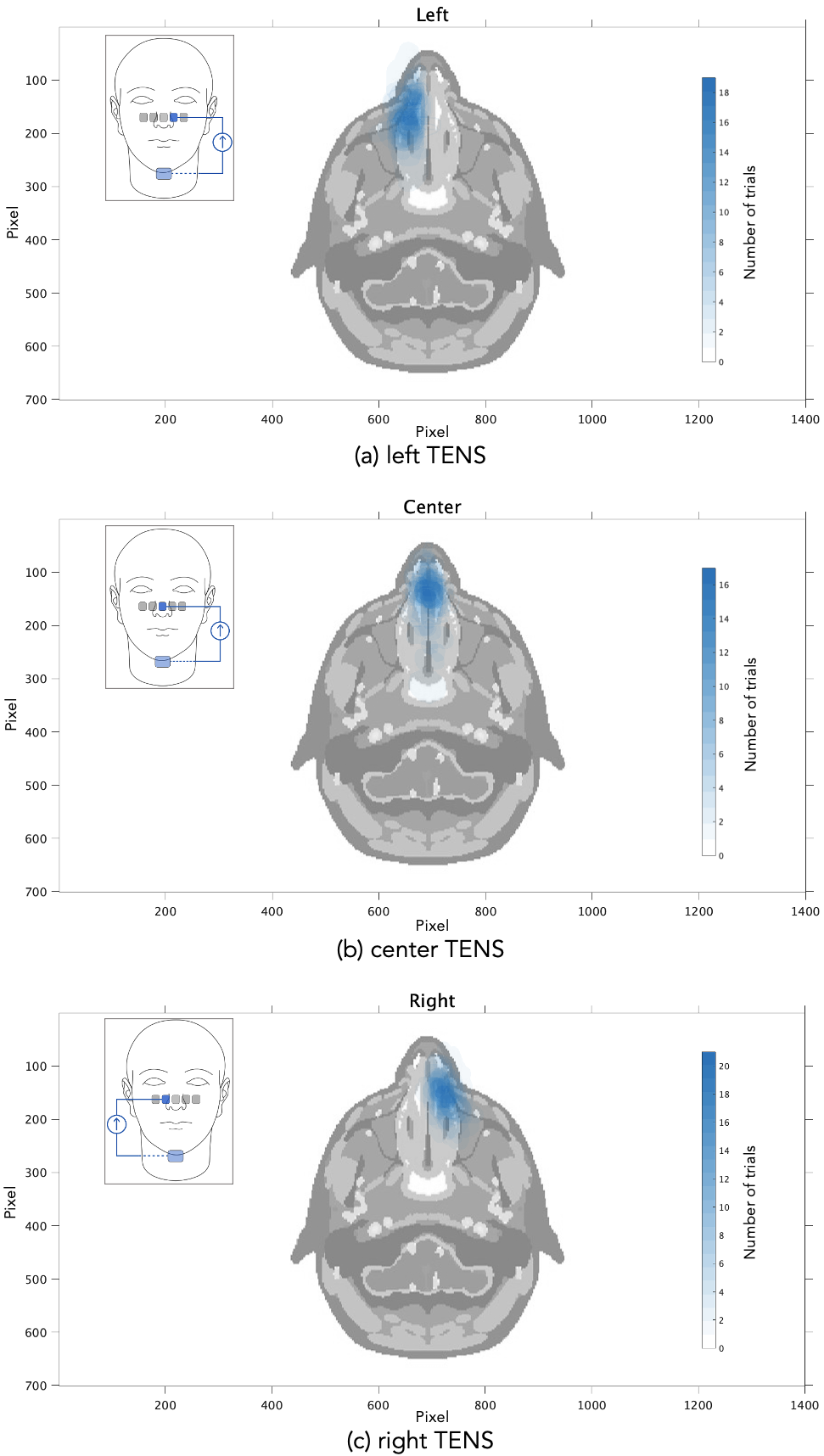}
 \caption{Areas where participants reported intranasal chemosensation with the cathode on (a) left side, (b) center, and (c) right side of the nasal bridge (Experiment 1). Among all valid trials, the areas drawn overlapped in at most (a) 19 trials, (b) 17 trials, and (c) 21 trials.}
 \label{fig:Ex1AreaDrawing}
\end{figure}

\begin{figure}[htbp]
 \centering
 \includegraphics[width=\columnwidth]{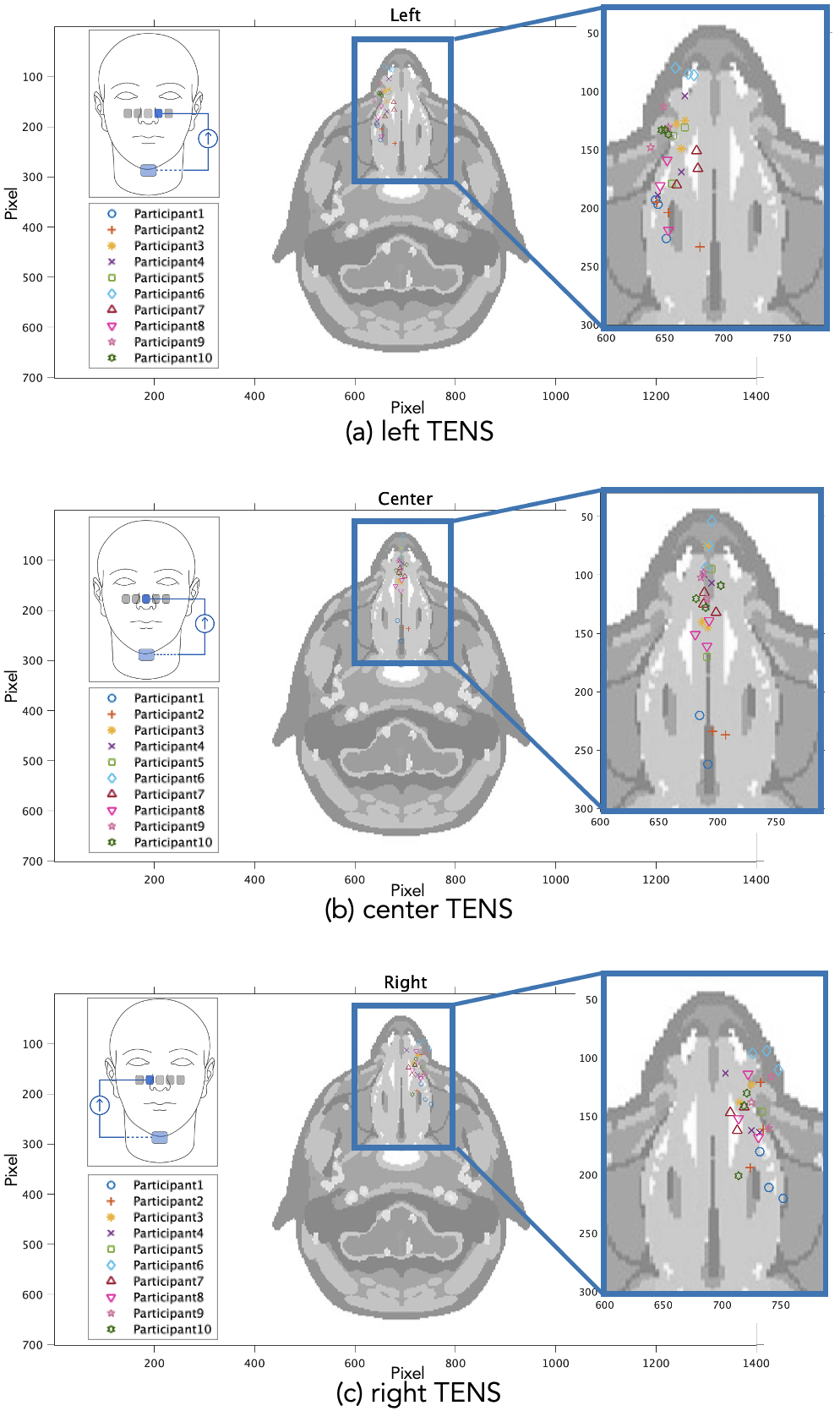}
 \caption{Points at which participants perceived intranasal chemosensation most intensely with the cathode on (a) left side, (b) center, and (c) right side of the nasal bridge (Experiment 1).}
 \label{fig:Ex1MaxPoints}
\end{figure}

\begin{figure}[htbp]
 \centering
 \includegraphics[width=\columnwidth]{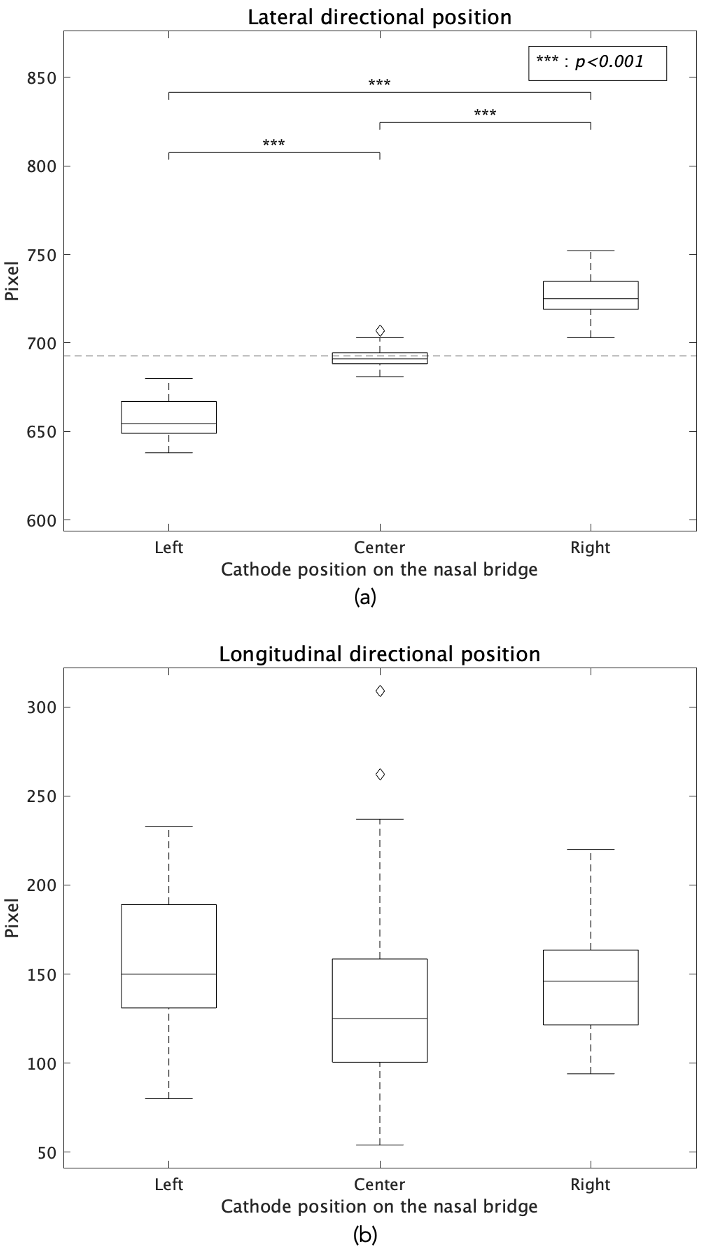}
 \caption{(a) Lateral and (b) longitudinal coordinates of the points of the highest intensity of intranasal chemosensation with the cathode on the left side, the center, and the right side of the nasal bridge (Experiment 1). The horizontal dotted line in (a) represents the lateral coordinates of the centerline of the head.}
 \label{fig:Ex1BoxPlots}
\end{figure}

Fig. \ref{fig:Ex1AreaDrawing} reveals the areas in which participants reported intranasal chemosensation, and Table \ref{tab:AreaTable} presents the mean and SD of the square measures of those areas under each condition.
The results summarized in Table \ref{tab:AreaTable} indicate that the position of the cathode on the nasal bridge does not affect the extent of the intranasal chemosensory area.
In Fig. \ref{fig:Ex1AreaDrawing}, the darker the blue color is, the more trials the participants marked the particular area in which they detected intranasal chemosensation.
Fig. \ref{fig:Ex1MaxPoints} reveals the points at which participants perceived intranasal chemosensation most intensely under each condition.
Statistical analyses using Kruskal--Wallis analysis of variance (ANOVA) revealed a significant difference between the lateral coordinates of these maximum intensity points ($p=1.054e-16$).
A post-hoc test using Scheffe's method of multiple comparison correction was performed, and statistically significant differences were found among all conditions, as displayed in Fig. \ref{fig:Ex1BoxPlots}(a) (Left-Center: $p=5.7969e-05$; Left--Right: $p=1.1451e-16$; Center-Right: $p=0.000279$).
The horizontal dotted line in Fig. \ref{fig:Ex1BoxPlots}(a) represents the lateral coordinates of the centerline of the head, which is 692.5.
By contrast, a Kruskal--Wallis ANOVA for the longitudinal coordinates of the maximum intensity points revealed no significant difference, as displayed in Fig. \ref{fig:Ex1BoxPlots}(b) ($p=0.15462$).

\begin{table}[htbp]
\caption{Mean and standard deviation of square measures of areas where participants reported intranasal chemosensation (Experiment 1).}
\label{tab:AreaTable}
\begin{tabular*}{80mm}{@{\extracolsep{\fill}}cccc}
                         & \multicolumn{3}{c}{Positions of the Cathodes on the Nasal Bridge (Pixel)} \\ \cline{2-4} 
                         & Left                   & Center               & Right\\ \hline
\multicolumn{1}{c}{mean} & 3506.35              & 3197.99            & 3603.59\\
\multicolumn{1}{c}{SD}   & 2328.12              & 2050.61            & 2256.30  
\end{tabular*}
\end{table}

\subsection{Second Experiment: Verification of the lateralization of intranasal chemosensation}
The corresponding rate, that is, how often the intranasal chemosensory perceived side matched the TENS-applied side was calculated for evaluation.
Table \ref{tab:correctRateTable} reveals the median and interquartile range (IQR) of the corresponding rate with TENS on the left, with TENS on the right, and all conditions summed.
A binomial test was performed on the number of corresponding answers without dividing the two conditions, and the null hypothesis that the corresponding rate is 0.5 was rejected ($p=5.871e-08$).

\begin{table}[htbp]
\caption{Mean, and interquartile range of the corresponding rate, which is how often the perceived side matched the TENS-applied side when the sham stimulus was applied opposite the TENS together (Experiment 2).}
\label{tab:correctRateTable}
\begin{tabular*}{80mm}{@{\extracolsep{\fill}}cccc}
                   & \multicolumn{3}{c}{TENS Stimulation Side} \\ \cline{2-4} 
                   & Left        & Right       & Total      \\ \hline
median             & 1           & 0.75        & 0.875        \\
IQR			& 0.25      & 0      & 0.125      
\end{tabular*}
\end{table}

\section{Discussion}
This study hypothesized that the lateralization of intranasal chemosensory perception is possible with the left TENS and right TENS, in which the electrodes on the nasal bridge of the TENS are shifted to the left and right.
The results of the two psychophysical experiments support this hypothesis.
Fig. \ref{fig:Ex1AreaDrawing}, \ref{fig:Ex1MaxPoints}, and \ref{fig:Ex1BoxPlots}(a) reveal that the location of intranasal chemosensation shifted significantly to the left by the left TENS and to the right by the right TENS, respectively.
Regarding the longitudinal direction, that is, the depth of the head, no significant difference in the intranasal chemosensory location among the left TENS, center TENS, and right TENS was observed.
The position of the electrodes on the nasal bridge only controls the lateral direction of intranasal chemosensory perception.
However, since TENS induces not only intranasal chemosensation but also tactile sensations on the skin surface and phosphenes, the location of intranasal chemosensory perception was possibly influenced by these sensations.
To verify that the lateralization of intranasal chemosensory perception was not caused by tactile sensation or phosphenes, Experiment 2 was conducted subsequently.

The results of Experiment 2, as summarized in Table \ref{tab:correctRateTable}, support the hypothesis that the lateralization of intranasal chemosensation is mediated by the distribution of stimuli to the trigeminal nerve branches.
Although the sham stimulation was less effective than the TENS in stimulating the trigeminal nerve, it provided tactile sensations on the cutaneous surface of the nasal bridge, as did the TENS.
Thus, if the tactile sensations cause the lateralization of the intranasal chemosensation, the rate at which participants perceive intranasal chemosensation from the side of the TENS should be approximately 0.5.
Despite that, participants answered the side of the TENS at a rate significantly higher than 0.5 as the location of the intranasal chemosensation. 
This result indicates that a factor other than tactile sensations causes lateralization.
In addition, as described in Section \ref{subsection:SimulationResults}, the difference in the phosphenes between the left and right eye was limited under the conditions of Experiment 2.
Regardless, the lateralization of intranasal chemosensory perception is evident, indicating that the lateralization of intranasal chemosensation is not caused by a left-to-right bias in the phosphenes.
Aoyama \textit{et al.}\cite{Aoyama2021} proposed the GOS (referred to as TENS in this paper) and argued that the method stimulates the olfactory nerve or trigeminal nerve because it induces intranasal chemosensation.
Based on the fact that numerous previous studies have proved that the distribution of trigeminal input is critical for the lateralization of intranasal chemosensation\cite{Kobal1989, Frasnelli2008, Kleemann2009, Lundstrom2012, Croy2014}, the trigeminal input was responsible for the lateralization observed in Experiment 2.
Aoyama \textit{et al.} did not reveal whether the TENS stimulates the olfactory nerves or the trigeminal nerves. 
However, the results of this experiment indicate that the TENS stimulates trigeminal nerves.
Additionally, participants described the intranasal chemosensation induced by TENS as ammonia-like, stinging, burning, and a pungent sensation as if water had been inhaled through the nose.
These descriptions are significantly similar to those of sensations induced by chemical substances known to stimulate the trigeminal nerve, such as CO$_2$ and ammonia, suggesting that TENS stimulates trigeminal nerves.

Some participants reported a ``metallic taste'' on the tongue along with a stinging chemical sensation in the nasal cavity.
Previous studies have shown that electrical stimulation of the mandibular nerve evokes a ``metallic taste''\cite{Nakamura2021, Lawless2005}. Accordingly, TENS possibly induces a slight current flow to the mandibular nerve; alternatively, electrical stimulation of the maxillary nerve possibly produces a slight metallic taste on the tongue.
The effect of the aforementioned ``metallic taste'' on the perception of chemosensation in the nasal cavity and a method to generate only intranasal chemosensation without metallic taste are possible avenues of future research.

This study demonstrated that the location of intranasal chemosensory perception could be manipulated by using an easy-to-wear method.
This result is highly applicable to VR and HCI.
For example, applications, such as locating the source of a pungent odor from its direction or navigation by chemosensation, are possible.

This method enables lateralization of intranasal chemosensation without blocking the nostrils.
Therefore, simultaneous chemical olfactory stimulation is possible while the TENS is applied.
In future studies, we plan to investigate the effects of simultaneous chemical odor presentation on olfactory perception.
An investigation into the perception of intranasal trigeminal stimulation when CO$_2$, a pure trigeminal stimulation, is presented simultaneously with TENS could also provide further insights into the mechanisms of intranasal chemosensory perception.
As reported by Aoyama \textit{et al.}\cite{Aoyama2021}, TENS can only induce irritating sensations, which limits its application to VR and HCI applications.
However, combining the TENS with chemical odor presentation in the future could considerably expand its range of applications.

\section{Conclusion}
This study demonstrated that intranasal chemosensation could be lateralized by shifting the electrode position of the nasal bridge to the left or right in the TENS which is used to induce an irritating sensation in the nasal cavity with electrodes on the nasal bridge and the back of the neck.
Finite element analysis simulations revealed that the left TENS and right TENS can selectively stimulate the left and right trigeminal nerve branches, which can lead to the hypothesis that the spatial perception of intranasal chemosensation, for which the distribution of stimuli to the trigeminal nerve branches is essential, could also be modulated.
Subsequent psychophysical experiments to test this hypothesis revealed that shifting the electrode on the nasal bridge to the left induced intranasal chemosensation on the left side in the nasal cavity, whereas shifting it to the right induced intranasal chemosensation on the right side.
Furthermore, the lateralization of intranasal chemosensation was not because of a perceptual illusion caused by the tactile sensation on the skin surface but by the distribution bias of stimuli to the trigeminal nerves.
The results of the study can contribute considerably to the academic field of chemosensory display techniques and have great potential for application to VR and HCI.

\bibliographystyle{ieeetr}

\begin{thebibliography}{99}
\bibitem{Aoyama2021}Aoyama, K., Miyamoto, N., Sakurai, S., Iizuka, H., Mizukami, M., Furukawa, M., Maeda, T. \& Ando, H. Electrical Generation of Intranasal Irritating Chemosensation. {\em IEEE Access}. \textit{9} pp. 106714-106724 (2021)
\bibitem{Archer2022}Archer, N., Bluff, A., Eddy, A., Nikhil, C., Hazell, N., Frank, D. \& Johnston, A. Odour enhances the sense of presence in a virtual reality environment. {\em Plos One}. \textit{17}, e0265039 (2022), http://dx.doi.org/10.1371/journal.pone.0265039
\bibitem{Nakamoto2018}Nakamoto, T., Ito, S., Kato, S. \& Qi, G. Multicomponent Olfactory Display Using Solenoid Valves and SAW Atomizer and its Blending-Capability Evaluation. {\em IEEE Sensors Journal}. \textit{18}, 5213-5218 (2018)
\bibitem{Holbrook2019}Holbrook, E., Puram, S., See, R., Tripp, A. \& Nair, D. Induction of smell through transethmoid electrical stimulation of the olfactory bulb. {\em International Forum Of Allergy \& Rhinology}. \textit{9}, 158-164 (2019), https://onlinelibrary.wiley.com/doi/abs/10.1002/alr.22237
\bibitem{Hariri2016}Hariri, S., Mustafa, N., Karunanayaka, K. \& Cheok, A. Electrical Stimulation of Olfactory Receptors for Digitizing Smell. {\em Proceedings Of The 2016 Workshop On Multimodal Virtual And Augmented Reality}. pp. 1-4 (2016), https://doi.org/10.1145/3001959.3001964
\bibitem{Kobal1989}Kobal, G., Van Toller, S. \& Hummel, T. Is there directional smelling?. {\em Experientia}. \textit{45}, 130-132 (1989), https://doi.org/10.1007/BF01954845
\bibitem{Croy2014}Croy, I., Schulz, M., Blumrich, A., Hummel, C., Gerber, J. \& Hummel, T. Human olfactory lateralization requires trigeminal activation. {\em NeuroImage}. \textit{98} pp. 289-295 (2014), https://www.sciencedirect.com/science/article/pii/S1053811914003693
\bibitem{Iannilli2008}Iannilli, E., Del Gratta, C., Gerber, J., Romani, G. \& Hummel, T. Trigeminal activation using chemical, electrical, and mechanical stimuli. {\em PAIN}. \textit{139}, 376-388 (2008), https://www.sciencedirect.com/science/article/pii/S030439590800242X
\bibitem{Kleemann2009}Kleemann, A., Albrecht, J., Schöpf, V., Haegler, K., Kopietz, R., Hempel, J., Linn, J., Flanagin, V., Fesl, G. \& Wiesmann, M. Trigeminal perception is necessary to localize odors. {\em Physiology \& Behavior}. \textit{97}, 401-405 (2009), https://www.sciencedirect.com/science/article/pii/S0031938409001243
\bibitem{Frasnelli2008}Frasnelli, J., Charbonneau, G., Collignon, O. \& Lepore, F. Odor Localization and Sniffing. {\em Chemical Senses}. \textit{34}, 139-144 (2008,11), https://doi.org/10.1093/chemse/bjn068
\bibitem{Hummel2002}Hummel, T. \& Livermore, A. Intranasal chemosensory function of the trigeminal nerve and aspects of its relation to olfaction. {\em Int Arch Occup Environ Health}. \textit{75} pp. 305-313 (2002)
\bibitem{kumar2012}Kumar, G., Juh\'asz, C., Sood, S. \& Asano, E. Olfactory hallucinations elicited by electrical stimulation via subdural electrodes: effects of direct stimulation of olfactory bulb and tract. {\em Epilepsy \& Behavior}. \textit{24}, 264-268 (2012)
\bibitem{Brooks2021}Brooks, J., Teng, S., Wen, J., Nith, R., Nishida, J. \& Lopes, P. Stereo-Smell via Electrical Trigeminal Stimulation. {\em Proceedings Of The 2021 CHI Conference On Human Factors In Computing Systems}. pp. 1-13 (2021), https://doi.org/10.1145/3411764.3445300
\bibitem{Nakamura2021}Nakamura, H., Mizukami, M. \& Aoyama, K. Method of Modifying Spatial Taste Location Through Multielectrode Galvanic Taste Stimulation. {\em IEEE Access}. \textit{9} pp. 47603-47614 (2021)
\bibitem{Laakso2013}Laakso, I. \& Hirata, A. Computational analysis shows why transcranial alternating current stimulation induces retinal phosphenes. {\em Journal Of Neural Engineering}. \textit{10}, 046009 (2013,7), https://dx.doi.org/10.1088/1741-2560/10/4/046009
\bibitem{Datta2009}Datta, A., Bansal, V., Diaz, J., Patel, J., Reato, D. \& Bikson, M. Gyri-precise head model of transcranial direct current stimulation: improved spatial focality using a ring electrode versus conventional rectangular pad. {\em Brain Stimulation}. \textit{2}, 201-207 (2009)
\bibitem{Baumann1997}Baumann, S., Wozny, D., Kelly, S. \& Meno, F. The electrical conductivity of human cerebrospinal fluid at body temperature. {\em IEEE Transactions On Biomedical Engineering}. \textit{44}, 220-223 (1997)
\bibitem{Gabriel2009}Gabriel, C., Peyman, A. \& Grant, E. Electrical conductivity of tissue at frequencies below 1 MHz. {\em Physics In Medicine \& Biology}. \textit{54}, 4863 (2009,7), https://dx.doi.org/10.1088/0031-9155/54/16/002
\bibitem{Gabriel1996}Gabriel, S., Lau, R. \& Gabriel, C. The dielectric properties of biological tissues: III. Parametric models for the dielectric spectrum of tissues. {\em Physics In Medicine \& Biology}. \textit{41}, 2271 (1996,11), https://dx.doi.org/10.1088/0031-9155/41/11/003
\bibitem{Lindenblatt2001}Lindenblatt, G. \& Silny, J. A model of the electrical volume conductor in the region of the eye in the ELF range. {\em Physics In Medicine \& Biology}. \textit{46}, 3051 (2001,10), https://dx.doi.org/10.1088/0031-9155/46/11/319
\bibitem{Hari1997}Hari, R., Portin, K., Kettenmann, B., Jousmäki, V. \& Kobal, G. Right-hemisphere preponderance of responses to painful CO2 stimulation of the human nasal mucosa. {\em PAIN}. \textit{72}, 145-151 (1997), https://www.sciencedirect.com/science/article/pii/S0304395997000237
\bibitem{Lundstrom2012}Lundstr\"om, J., Gordon, A., Wise, P. \& Frasnelli, J. Individual Differences in the Chemical Senses: Is There a Common Sensitivity?. {\em Chemical Senses}. \textit{37}, 371-378 (2012,1), https://doi.org/10.1093/chemse/bjr114
\bibitem{Moessnang2011}Moessnang, C., Finkelmeyer, A., Vossen, A. \& Schneider, U. Assessing Implicit Odor Localization in Humans Using a Cross-Modal Spatial Cueing Paradigm. {\em PLOS ONE}. \textit{6}, 1-11 (2011,12), https://doi.org/10.1371/journal.pone.0029614
\bibitem{Wu2020}Wu, Y., Chen, K., Ye, Y., Zhang, T. \& Wen Zhou Humans navigate with stereo olfaction. {\em Proceedings Of The National Academy Of Sciences}. \textit{117}, 16065-16071 (2020)
\bibitem{Hasgall2018}Hasgall, P., Di Gennaro, F., Baumgartner, C., Neufeld, E., Lloyd, B., Gosselin, M., Payne, D., Klingenböck, A. \& Kuster, N. IT'IS database for thermal and electromagnetic parameters of biological tissues. {\em Version 4.0}. (2018)
\bibitem{Maharjan2018}Maharjan, A., Wang, E., Peng, M. \& Cakmak, Y. Improvement of Olfactory Function With High Frequency Non-invasive Auricular Electrostimulation in Healthy Humans. {\em Frontiers In Neuroscience}. \textit{12} (2018)
\bibitem{Cakmak2020}Cakmak, Y., Nazim, K., Thomas, C. \& Datta, A. Optimized Electrode Placements for Non-invasive Electrical Stimulation of the Olfactory Bulb and Olfactory Mucosa. {\em Frontiers In Neuroscience}. \textit{14} (2020)
\bibitem{Badran2022}Badran, B., Gruber, E., O'Leary, G., Austelle, C., Huffman, S., Kahn, A., McTeague, L., Uhde, T. \& Cortese, B. Electrical stimulation of the trigeminal nerve improves olfaction in healthy individuals: A randomized, double-blind, sham-controlled trial. {\em Brain Stimulation}. \textit{15}, 761-768 (2022), https://www.sciencedirect.com/science/article/pii/S1935861X22000845
\bibitem{Negoias2013}Negoias, S., Aszmann, O., Croy, I. \& Hummel, T. Localization of Odors Can Be Learned. {\em Chemical Senses}. \textit{38}, 553-562 (2013,6), https://doi.org/10.1093/chemse/bjt026
\bibitem{Lawless2005}Lawless, H., Stevens, D., Chapman, K. \& Kurtz, A. Metallic Taste from Electrical and Chemical Stimulation. {\em Chemical Senses}. \textit{30}, 185-194 (2005,3), https://doi.org/10.1093/chemse/bji014
\end{thebibliography}

\begin{IEEEbiography}[{\includegraphics[width=1in,height=1.25in,clip,keepaspectratio]{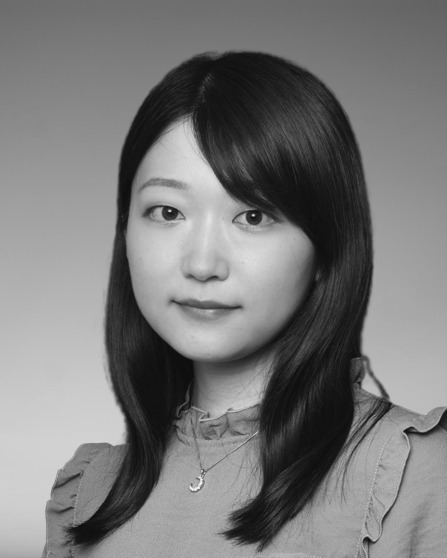}}]{Ayari Matsui} received the B.E. and M.E. degrees in mechano-informatics from the University of Tokyo in 2021 and 2023, respectively. She currently works as a data scientist at a leading IT company in Japan. Her research interests include virtual reality, olfactory interfaces, and multi-modal interfaces.
\end{IEEEbiography}

\begin{IEEEbiography}[{\includegraphics[width=1in,height=1.25in,clip,keepaspectratio]{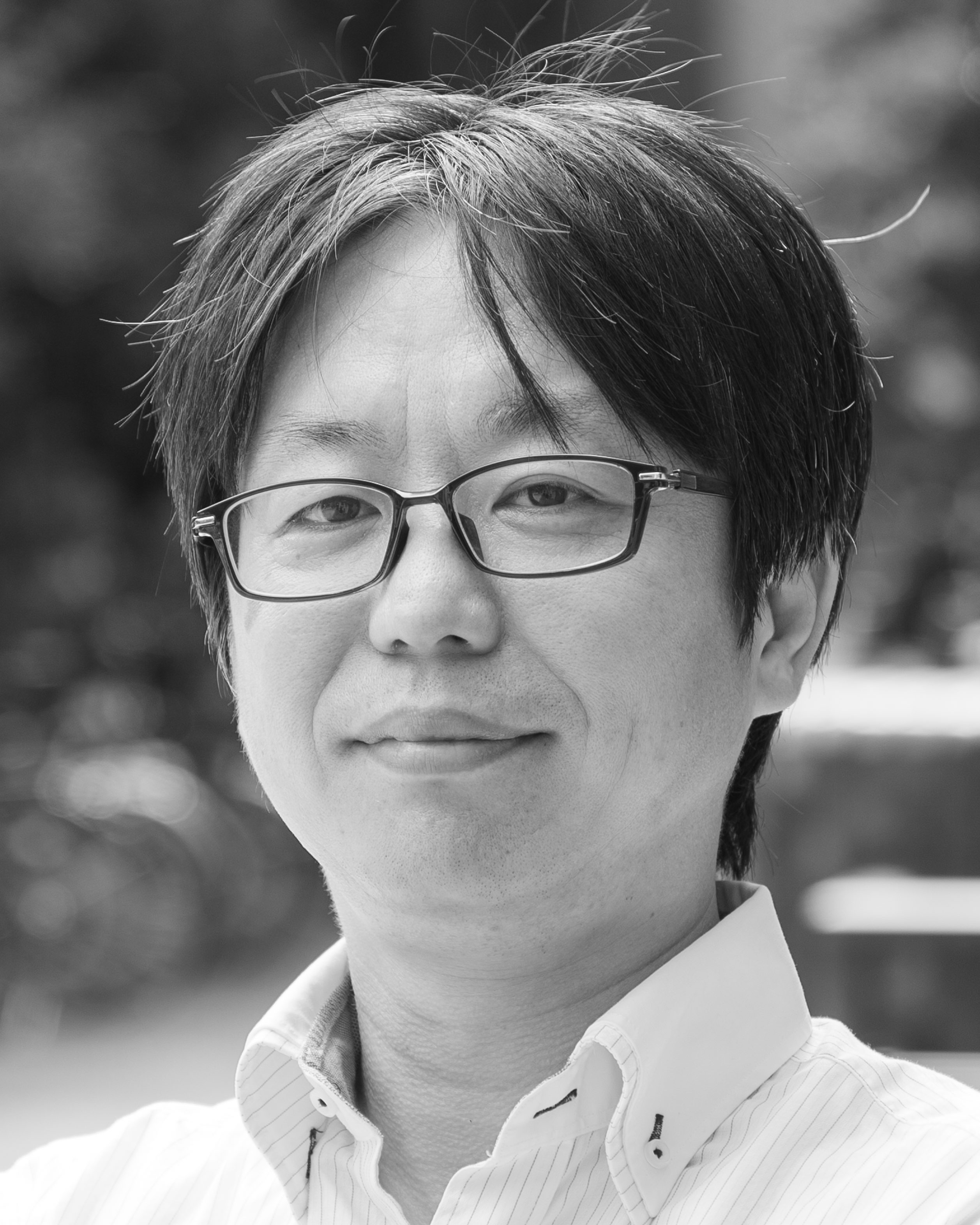}}]{Tomohiro Amemiya}(M'08) received the B.S. and M.S. degrees in mechano-informatics from the University of Tokyo in 2002 and 2004, respectively, and the Ph.D. degree in bio-informatics from Osaka University in 2008. From 2004 to 2019, he was a Researcher with the NTT Communication Science Laboratories. He was also an Honorary Research Associate with the Institute of Cognitive Neuroscience, University College London, from 2014 to 2015. Since 2023, he is a Professor with the Information Technology Center and the Virtual Reality Educational Research Center at the University of Tokyo. His research interests include haptic perception, tactile neural systems, and human--computer interaction using sensory illusion and virtual reality technologies. Dr. Amemiya has received several awards, including the Grand Prix du Jury (Laval Virtual International Awards 2007) and Best Demonstration Award (Eurohaptics 2014).
\end{IEEEbiography}

\begin{IEEEbiography}[{\includegraphics[width=1in,height=1.25in,clip,keepaspectratio]{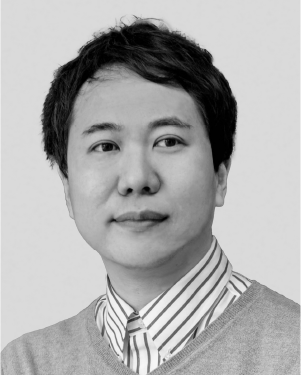}}]{Takuji Narumi} received the B.E. and M.E. degrees from the University of Tokyo in 2006 and 2008, respectively, and the Ph.D. degree in engineering from the University of Tokyo in 2011. He is an Associate Professor with the Graduate School of Information Science and Technology, the University of Tokyo. His research interests include the intersection of technologies and human science. He has been researching extending human senses, cognition, and behavior by combining virtual reality and augmented reality technologies with findings from psychology and cognitive science. Dr. Narumi was a recipient of several awards, including MEXT The Young Scientists' Award, SIGCHI Japan Chapter Distinguished Young Researcher Award, Japan Media Arts Festival 2017 Excellence Award, and CHI Honorable Mentions (2019, 2020).
\end{IEEEbiography}

\begin{IEEEbiography}[{\includegraphics[width=1in,height=1.25in,clip,keepaspectratio]{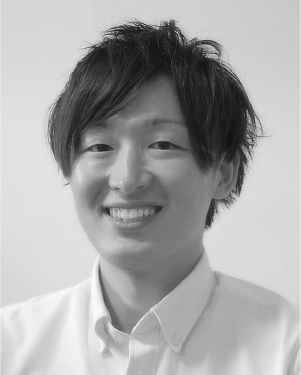}}]{Kazuma Aoyama} received the Ph.D. degree in information science from Osaka University in 2016. Since 2023, he has been a Associate Professor at Gunma University and Project Associate Professor with the Virtual Reality Educational Research Center at the University of Tokyo. His current interests include virtual reality, brain science, and neuro-engineering.
\end{IEEEbiography}

\EOD

\end{document}